# The Viscosity of Liquids in the Dual Model.

*Fabio Peluso*

*Leonardo SpA – Electronics Division*
*Via Monterusciello 75, 80078 Pozzuoli (NA) - Italy*
*mailto: fpeluso65@gmail.com.*

## Abstract

In this paper, we develop a reliable model of the viscosity in liquids in a liquid dual model framework. The analytical expression arrived at exhibits the correct $T$ – dependence Arrhenius-like exponential decreasing trend; it is compared with the experimental values of viscosity for water with acceptable accuracy, and those related to the mechano-thermal effect in liquids under low-frequency shear, for which the first-ever theoretical interpretation is given, supported by agreement with experimental data. It is even shown that a numerical approach of viscosity in liquids dealt with as dual systems, provides good agreement with experimental data. The expression of viscosity shows an explicit dependence upon the sound velocity and the collective vibratory degrees of freedom (DoF) excited at a given temperature. At the same time, the terms involved depend upon the Boltzmann and Planck constants. Finally, the physical model is coherent with the postulate of microscopic reversibility as well as with the time's arrow for macroscopic dissipative mechanisms.

## 1. Introduction.

One of the challenges, probably the most relevant, of the condensed matter physics is the understanding of "macroscopic" phenomena through the modeling at "mesoscopic" scale. Among such phenomena, we cite the thermal conductivity, the specific heat, the propagation of energy and momentum under the action of external stimuli, etc. Undoubtedly, the viscosity of liquids is among these phenomena, in fact it plays an important role in both scientific theory and daily life. Therefore, it can be understood that the models of viscosity are evaluated from various points of view by the experts in different fields. On one hand, industrial experts are mainly interested in a macroscopic description of the viscosity of liquids. On the other hand, physicists are interested in how the properties of the viscosity depend on the molecular structure of the liquid and on the properties of the individual molecules known from other phenomena or from other correlations. Understanding the physical mechanisms at the base of the viscosity in liquids is in fact crucial even to solve the conundrum of some microscopic physical mechanisms yet unclear today, such as for instance the mechano-thermal effect in liquids under low-frequency shear revealed by Noirez and co-workers [1-9], the viscosity minimum and the dependence of viscosity upon the speed of sound and the collective DoF [10-11], as well as the superfluidity of HeII [12-16].

The physical modelling of the viscosity, especially in liquids, is a topic to which physicists devoted in the past a desultory attention, while the advent of new liquid modelling inspired by recent experimental findings, has revived such topic. The first who formally put on the table the problem was Newton; in his famous *Philosophiae Naturalis Principia Mathematica* although he did *not* report the equation that defines the viscosity $\eta$ as the quantity relating the shear force per unit area, $f_{x,y}$, to the shear gradient, $\frac{d\upsilon_x}{dy}$,

1. $f_{x,y} = -\eta \frac{d\upsilon_x}{dy}$,

the germ of the idea is there [17]. This equation is often called *Newton's law of viscosity*; it states that the shearing force per unit area is proportional to the negative of the shear gradient. Actually, Newton suggested it as an empiricism more than a law, i.e. the simplest way for relating the shear gradient and the stress. Nevertheless, experiments demonstrate that it is valid for all fluids with molecular weight of less than about 5000, usually referred to as *Newtonian* fluids. As far as polymeric liquids, suspensions, pastes, slurries, and other complex fluids, as also the blood for instance, their resistance-to-flow behaviour is much more complex than Eq. (1), because the viscosity is no longer independent of the velocity gradient, so that they are referred to as *non-Newtonian* fluids. Equation (1) provides the relation one should use in calculating the dissipation process at stake when fluids are in movement or when a solid body moves through a fluid. The next step for a physicist is of course that of stating a physical model at the atomic/molecular level for the viscosity, and to reproduce its dependence upon the fundamental variables.

As is well known, the viscosity of gases was modelled by Maxwell based on a pure molecular kinetic approach [18]. The physical mechanism at the base of the viscosity in gases is the momentum exchange between molecules, supposed of course free from any interaction due to the large intermolecular distance, only contact forces being allowed. On the basis of the theory, further developed by Chapman and Enskog [19-23], Hirschfelder [24] and Grad [25-27], finally reworked by Cohen [28-32], the calculated values of the constants in the equation of viscosity are in good agreement with the experimental data. The viscosity of gases can be calculated from the properties of their atoms, or molecules; interestingly, Lennard-Jones [33] deduced the law of forces acting between different atoms from the measured values of the viscosity of gases. The expression arrived at by Maxwell accurately reproduces also the temperature dependence of the viscosity of gases $\eta_g$ upon absolute temperature $T$ as proportional to its square root, $\eta_g = \frac{2}{3\pi^2 d^2}\sqrt{\pi m_g K_B T}$, where $m_g$ is the average molecular mass, $d$ the average distance of the molecules and $K_B$ the Boltzmann constant. The reason is simply that the momentum exchanged between molecules increases with temperature, because the same does their average velocity.

The situation with liquids is completely different for several reasons, of which we list hereafter only the main ones. Firstly, unlike gas molecules, liquid molecules interact among them by means of intense forces, comparable to those among solid molecules, and working at distance; therefore, the presence of an interaction potential prevents to deal liquid molecules as hard spheres. Secondly, unlike solid molecules, liquid molecules not only oscillate around equilibrium positions, but also diffuse over distances comparable to, if not larger than, their size, similarly to the molecules of a gas. In short, they combine the two types of motions, of solid and gas molecules, making very challenging the feasibility of a general theory of liquids. These considerations were condensed in a simple concept by Landau, "Liquids do not have small parameters", [34]. Thirdly, the dependence of the viscosity of liquids upon absolute temperature is completely different from that of gases and cannot be reproduced on a theoretical scheme based on a kinetic approach, as Maxwell did for the gas. In fact, while the viscosity of gases $\eta_g$ increases with temperature, that of Newtonian liquids $\eta_l$ is experimentally found to exponentially decrease with $T$,

2. $\eta_l(T) = A e^{-\frac{b}{T}}$

where $A$ and $b$ are constants depending on the liquid nature and slowly dependent upon temperature. S. A. Arrhenius proposed for the first time in 1887 the same law for the internal friction of dilute solutions [35] based on experimental data; two years later, he proposed the same type of equation for the dependence upon temperature of the velocity constant in chemical reactions [36-37]. The physical reason why an expression such as Eq.(2) holds for the chemical reactions is that the concentration of reagents decreases with temperature (and with time, while the reaction is in progress), but the same reason cannot be advocated for the liquid molecules in fluids in movement. Therefore, theorists were compelled to adopt physical mechanisms different from the kinetic approach for the viscosity of liquids, with the additional constraint of reproducing an Arrhenius behaviour.

As we shall see in detail in the next section, the viscosity modelling of liquids was generally approached from the macroscopic – phenomenological point of view, i.e. without any connection with the real world of liquid' constituents, the atoms or molecules (with the exception of the Eyring' model). Being conscious that the pure gas-kinetic approach could not be successful if applied to liquids, Maxwell used the relaxation time $\tau$ to provide a mathematical formulation of the problem in terms of viscoelasticity. The model of Maxwell, the historically most important among the phenomenological ones, is a mathematical interpolation between a pure elastic solid response and that of a pure dissipative fluid. The Maxwell analogue of liquids is in fact represented by an elastic element, i.e. a Hookean spring, in series with a dissipative element, a damper. It represents the most simple model for viscoelastic materials, typical of liquids: it provides viscous flow on long timescale, and additional elastic resistance to fast deformations (the opposite behaviour, i.e. of fluids showing a viscous response on short timescale and an elastic response on long ones, is given by the Voigt-Kelvin analogue, where the two elements, the elastic spring and the damper, are in parallel instead than in series [38-39]).

The topic of the present manuscript is to show that a reliable liquid viscosity modelling can be framed into the Dual Model of Liquids [40-45] (DML), at least for the contribution due to the presence of collective modes that affects the behaviour of liquids at mesoscopic scale. Throughout the manuscript it is supposed to be in presence of laminar flow, as described in Figure 1, and that the coupling between the resistance force and the gradient of velocity along the normal direction is represented by Eq.(1) (for a generalization of Eq.(1) to the 3D geometry, see for instance [46], section 1.2).

The structure of the paper is the following: in Section 2 we will make a brief historical overview of the main theoretical approaches for the viscosity of liquids. The model of liquid viscosity is described in Section 3, and the comparison with experimental data is given in section 4. The Section 5 is dedicated to discussing the related details. Finally, the Conclusions is dedicated to propose experiments aimed at verifying the model illustrated in Section 3, as well as other possible applications and capabilities of the DML.

## 2. The Main Theoretical Advances in the Modelling of the Viscosity of Liquids.

This paragraph has the dual purpose of showing the limits of the main historically developed models for the viscosity of liquids, and the similarities of the basic hypotheses of some of them with those of the DML.

The theoretical modelling of the viscosity of liquids $\eta_l$ was faced by several authors in the past century, based on as much different approaches. Being equation (2) of phenomenological origin, the goal of any physical model of viscosity in liquids is to reproduce it as functional dependence upon the absolute temperature. Equation (2) was proposed also by Andrade in the 30's of the past century [47-50] on a pure empirical base, in the attempt to reproduce the experimental data on liquid viscosity and to compile the several theoretical approaches known at that time. As Andrade himself highlights [47], there were many empirical formulas proposed until then due to Poiseuille, van der Waals jr, Koch, Meyer, Slotte, Frenkel, Fleming, Madge, Black [48], and others, connecting viscosity with temperature. However, they all had at least three arbitrary constants and a limited range of validity. Moreover, some of them did not reproduce correctly the temperature dependence of $\eta_l$. Finally, different types of formula were used depending on the different types of liquids. Equation (2) proposed by Andrade was instead valid for all the types of liquids, the accommodation depending only on the different values attributed to the two variables, $A$ and $b$, whose origin and physical meaning still remained unclear at that time.

In the past century, Kirkwood and co-workers [51-53] developed a rigorous kinetic theory of the transport properties of monatomic liquids. However, this theory did not lead to easy-to-use results. An older theory, developed by Eyring and co-workers [54-66], although less well-grounded theoretically, did give a qualitative picture of the mechanism of momentum transport in liquids, permitting rough estimation of the viscosity from other physical properties. In the past century many other minor models of viscosity were developed, however, because there are many textbooks where they are described [46,67-68], here in this paper we will not report their details. Nevertheless, we will recall the reader' attention on the two models that more closely reproduce the dynamics characterizing the viscosity at molecular level, namely the Reaction Rate Theory by Eyring and the qualitative approach of Andrade. The reader will see by himself how these two approaches share their basic hypotheses with the DML and with the ideas of Frenkel.

To reproduce the empirical law given by Eq.(2), Eyring and Hirschfelder imagined a physical process similar to that of the self-diffusion. In the frame of the Eyring model, the individual molecules of a pure liquid at rest, although constantly in motion, are pictured as vibrating around an equilibrium position within a "cage" formed by its nearest neighbors, because of their close packing. This cage is represented by an energy barrier of height $\Delta G^+$, where $\Delta G^+$ is the free energy of activation for escaping from the cage in the stationary fluid. The molecules vibrate in their cages until a) they acquire sufficient energy to overtake the attracting forces "and" b) a hole is available in the nearest neighbour where the molecule can jump and stay there for a time sufficient to re-establish the Maxwellian velocity distribution (see also Figure 2 in [41]). This game recalls in some way that of the electrons in a semiconductor, giving to the Eyring model of the liquid a solid-like character. It was used by Eyring and Hirschfelder to justify the translational movement of molecules in liquids, in such a way that they could transport momentum in their cage-to-hole wandering. This mechanism differs from that of self-diffusion only for the fact that the motion is here driven by the velocity gradient, instead of being random. The molecules thus move in each of the coordinate directions in jumps of length $a$ at a frequency $\nu$ per molecule. The rate equation

3. $h\nu = K_B T \exp\{-\Delta G^+/K_B T\}$,

where $h$ is the Planck constant, allows to obtain the frequency of the molecule jumps.

It is worth highlighting two facts, namely i) that a similar equation was set up by Eyring a few years before to describe the rate of a chemical reaction as function of the temperature (as S. Arrhenius did in 1889, [36,37]), and ii) that this frame was not completely new yet at the time when Eyring introduced it. As for the first point, the Eyring equation in chemical kinetics, also known as the Rate Reaction Equation, is an equation that describes the rate of reaction as a function of temperature. The equation derives from the transition state theory which, unlike the empirical Arrhenius equation, is a theoretical model based on statistical thermodynamics. It was developed by Michael Polanyi, Meredith Gwynne Evans and Henry Eyring almost simultaneously in 1935 [69-73]; it was presented in a form similar to Eq.(3), where $\Delta G^+$ was interpreted as the Gibbs free energy of activation for the chemical reaction. A few years later, Eyring and Hirschfelder adapted that expression for the energy linked to the molecule' cage-to-hole jumps, giving to the viscosity model the name of "Rate Reaction Theory", loaned from its former introduction in chemistry, later named also the "Hole Theory". As it concerns the second point, the idea that liquids are made by a very large number of randomly oriented (ephemeral) pseudo-crystals of submicroscopic size, that continuously rearrange their position and composition in terms of atoms or molecules, was introduced by Stewart around 30's, and reported later by Frenkel [74]. Stewart proposed to dub as "cybotactic groups" such submicroscopic crystals, consisting of few tens of molecules, and assumed them to be connected with each other by thin layers of the wholly amorphous phase. This idea is similar to that advanced by Andrade [47,49-50]. Frenkel used the relaxation time $\tau_F$, introduced for the first time by Maxwell [75], as the average time between particle jumps at one point in space in a liquid; its inverse, $\nu_F = 1/\tau_F$, is the frequency of occurrence of the particle jumps. As we will see, a similar idea for the continuously rearranging pseudo-crystals, was adopted also by Andrade. We have further elaborated this concept in the DML, see Figure 2 in [41], where we introduced the concept of "wandering icebergs" jumping between two equilibrium positions in a liquid, replacing therefore a single molecule with a solid-like iceberg. We will return on this and other related points in the Discussion.

Returning to the Eyring-Hirschfelder model, they provided the distribution function of the molecules in terms of the heat of evaporation, supposed to be equal to the energy barrier $\Delta G^+$ that the molecules have to overcome to displace relatively to each another, thus generating the viscous flow. As the authors highlight in their manuscripts, the rate of this process is determined by factors similar to those acting in the chemical reactions, therefore they treated the viscous flow on a base similar to that of the Reaction Rate theory. Within this frame, Eyring and Hirschfelder got the expression

4. $\quad \eta_l = \dfrac{hN}{V} \exp\left\{\dfrac{T_b}{T}\right\}$,

with $T_b$ the boiling temperature, for the dynamic viscosity of simple monatomic liquids in agreement with the long-used and successful empiricism given by Eq.(2) [45].

The Eyring theory is an approximate description of the phenomena of viscous flow even for liquids of simple structure. The basic equations do not contain any data related to the molecular potential field and the structure of the liquid, that should be estimated empirically to apply the theory in actual cases. A further weak point of this model is that no special term regarding the holes required for the displacement of molecules is involved [45]. McLaughlin further elaborated the theory of rate processes to fill this deficiency [76].

The approach followed by Andrade is based on completely different basic hypothesis. He presented his ideas in two steps, the first in 1930 in a paper on Nature [47], to which many illustrious authors replied; the replica by Andrade came few weeks later [48]. However, curiously only 4 years later he presented on another magazine his model of viscosity of liquids [49-50], previously announced, in a better-structured frame. In his pioneering works Andrade proposed the ingenious idea that the viscosity in flowing liquids is due to the capability of molecules belonging to contiguous layers to make temporary bindings, whose duration not exceeding the time required by the molecules to acquire a common velocity of translation. The union takes place under the action of the residual field that suffices to bind them temporarily in the solid-like state, while the thermal unrest ensures its limited duration. Historically he was the first, even before Brillouin [77] and Frenkel [74], to suppose that liquids have a double soul, advancing the hypothesis that "liquids crystallize (locally) in minute patches: at the temperature of solidification the crystallization becomes general and permanent, while at the boiling point it must be very small. From here the tendency of molecules to combine in large, ephemeral, structures, opposed by the thermal unrest." [48]. It is then clear how the greater the temperature, the lesser the probability of forming the temporary solid-like structures. Consequently, the viscosity of liquids would decrease with rising the temperature until the energy of motion overcomes the molecular residual field. The Andrade model of liquid structure and viscosity is not supported by a theoretical building, to the point that he doesn't provide a way of how to construct an expression for $\eta_l$. He himself points out that this is the same frame on which Maxwell based his theory of viscosity for gases; nevertheless, because the Maxwell approach would have led to a viscosity increasing with temperature rather than decreasing, a different picture would have been needed for liquids [47-49].

The thermodynamic approach of Nettleton [78] to modelling the viscoelasticity in liquids at high frequencies deserves to be discussed. He proposed a tensorial ordering parameter, obeying a linear relaxation equation, to describe the elastic shearing strain exhibited by a liquid at high frequencies. Very interestingly, he shows that the shear modulus in the stress tensor and the coefficients of the relaxation equation are linked via the Onsager-Casimir reciprocal relations. This fact implies that the total rate of deformation is divided in two parts, the elastic and the viscous contributions, in agreement with the hypothesis made by Frenkel when he derived the viscoelastic relaxation equation of Maxwell. Equally interesting is that even Nettelton uses the concept of cybotactic groups of molecules for his liquid model; the groups are regarded as regular lattices, although alternated with holes to permit molecules to jump from one place to an adjacent hole. The quoted paper, however, does not contain any comparison with existing experimental data.

Jumping to more recent years, Zaccone has proposed an interesting general theory of the viscosity of liquids based on Non Affine particle motions [79-80]. He obtained an expression for the viscosity of liquids and solids derived from the Hamiltonian for atomistic motion in the framework of the so-called Non Affine linear response theory. The expression arrived at provides a direct connection of the viscosity with the speed of sound and the Vibrational Density of States (VDOS). We will return on this aspect in the Discussion to show that the Zaccone' expression and that derived in this manuscript in the DML framework share the same characteristic, i.e. the dependence on the speed of sound and on the collective DoF. The Zaccone' expression, however, lacks of a direct comparison either with experimental data or the Arrhenius behaviour.

Modern techniques used to compute the viscosity of liquids (and solids) from first principles are based on the Green-Kubo formalism that is among the most successful methods. Such techniques, however, have the week point that they do not lead to compact and closed-form expressions; besides, they are employed for numerical calculations of viscosity, in turn based on previous molecular dynamics simulations.

In the next section, we will build the model of viscosity in the DML framework and show that it characterized by an Arrhenius trend.

## 3. The theory of the Viscosity of Liquids in the DML

### *3.1    The Physical Model.*

The dual model that we are going to use to construct the model of viscosity in liquids, DML [41], rests on two fundamental hypotheses, both having an experimental background [81-100]. The first assumes that the liquids are organized at mesoscopic scale by means of solid-like structures, whose number and size, of a few molecular diameters, varies with the temperature and pressure of the liquid. Within such clusters, the elastic waves propagate as in the corresponding solid phase. The second hypothesis is simply that the two subsystems interact among them, allowing to explain energy, momentum and mass diffusion in liquids [41-45]. The dualism of liquids is a recurring topic in physics, not only in modelling the structure of "normal" liquids. One should remember that Landau too described the HeII as made of "normal" and "superfluid" parts [12]). Similar ideas were advanced in the past century by several scientists, in particular by those who put down the basis for the first models of the viscosity of liquids. Without neglecting the relevant contributions by Debye, Brillouin and Frenkel, as well Eyring and Andrade, more recently Fiks used them again in the early '60s to explain thermal diffusion in liquids [104] and the electronic drag and thermal diffusion in metals [105], while Andreev [106], assuming that a liquid consists of two weakly coupled systems, the phonons and the remainder of the liquid, calculates the effects connected with the presence of weakly damped phonons in a normal liquid. Much more recently, in the Phonon Theory of Liquid Thermodynamics  (PLT), [107-117], the authors advanced the same hypothesis on the base of which they deduced for the first time an expression for the specific heat valid in gases, liquids and solids. One of the direct consequences of the above picture is that what we usually define "liquid" is not meant in the DML as a liquid following the classical definition, but a mixture of solid icebergs and an amorphous phase (i.e. a classical liquid). The lifetime of such solid-like structures is very small, and comparable with the relaxation times characterizing the transport processes in liquids. About this point, very interesting are the results of recent experiments performed in water and glycerol [100], that revealed for the first time the presence and long-lasting permanence of elasticity on a mesoscopic scale. The elastic response in such liquids persisted indeed for time laps of microseconds, hence four orders of magnitude longer than the typical intermolecular relaxation time. Such rubber-like elasticity and large strain response in fluid glycerol were interpreted by the authors as due to the relaxation processes of collective modes in metastable groups of molecules, those we dub icebergs, or *liquid particles*, in the DML. This sounds like a recurrent idea similar to those of Andrade, Stewart, Frenkel or Brillouin, and finally adopted in the DML, although in a modified dress. Consequently, one may argue that the lifetime of such icebergs is equal to, if not larger than, the relaxation time of elastic propagation. The authors also point out that such modes require the existence of a transient state with solid-like long-range correlations, although different from the bulk state [41].

More specifically, in the DML, we are used to distinguish between the *liquid particles*, i.e. the solid-like aggregates of molecules, from the amorphous phase, typical of liquids, where molecules fluctuate independently from each other. In the same way, we distinguish between the *lattice particles*, i.e. the harmonic quantized waves crossing the liquid particles, and the wave-packets, i.e. the anharmonic waves in which *lattice particles* commute when they interact with the *liquid particle* borders (see Figure 1 in [41]). As far as the perturbations travel inside a solid-like structure, they show a solid-like character, i.e. they are (quasi) harmonic waves travelling at speeds close to that of the corresponding solid phase (3200 m/s in the case of water, [78,41]). When they leave the *liquid particle* and travel through the amorphous environment, they lose the harmonic outfit and generate wave-packets. The interaction itself between these waves and the solid-like icebergs have an anharmonic character, allowing momentum and energy be exchanged between the two populations of pseudo-particles, the *liquid particles* and the *lattice particles*. Figure 2 shows way how *lattice* and *liquid particles* interact among them, giving rise to the energy, mass and momentum transport in liquids. The free-flight of a wave-packet between two successive interactions with as many *liquid particles*, during which they may be considered as local microscopic heat currents, is long $\langle \Lambda_{wp} \rangle$ and lasts $\langle \tau_{wp} \rangle$. Obviously, $\langle \nu_{wp} \rangle = 1/\langle \tau_{wp} \rangle$ is the average frequency of the elementary interactions represented in Figure 2. Considering a surface S arbitrarily oriented within the liquid, at thermal equilibrium an equal number of wave-packets will flow through it in opposite directions. Such heat currents are driven locally by a virtual temperature gradient $\langle \frac{\delta T}{\delta x} \rangle$ applied over $\langle \Lambda_{wp} \rangle$, keeping anyway their distribution in the liquid isotropic. If $q_T$ is the total heat content of the liquid, the fraction associated with the wave packets is

$$q_T^{wp} = m q_T = m \frac{\int_0^T \rho C_V d\theta}{\rho C_V T} \rho C_V T = m^* \rho C_V T$$, where $C_V$ is the specific heat at constant volume, , $\rho$

the medium density (the readers interested to the formal building of the model, are warmly addressed to the related literature [41]). The variation of the energy density within the liquid along the "+*x*" direction is:

5. $\delta_{+x} q_T^{wp} = \frac{1}{6} \delta(m q_T) = \frac{1}{6} \frac{\partial q_T^{wp}}{\partial T} \delta T = \frac{1}{6} \frac{\partial q_T^{wp}}{\partial T} \left\langle \frac{\delta T}{\delta x} \right\rangle_{+x} \left\langle \Lambda_{wp} \right\rangle$

where the local virtual temperature gradient $\left\langle \frac{\delta T}{\delta x} \right\rangle$ takes the role of the thermodynamic "generalized" force [with the meaning of Non Equilibrium Thermodynamics (NET)] driving the diffusion of thermal excitations along $x$. The factor $1/6$ accounts for the isotropy of the phonon current through a liquid at equilibrium. The parameter $m$ is the ratio between the number of collective DoF surviving at temperature $T$ and the total number of available collective DoF for the system, then it holds $0 \leq m \leq 1$. This parameter plays a pivotal role in the DML; in the Discussion we will show its close relation with the Vibratory Density of States (VDOS) [41,118]. By applying this driving force over the distance $\left\langle \Lambda_{wp} \right\rangle$, a wave-packet diffusion, i.e. the heat current:

6. $j_{+x}^{wp} = -D_+^{wp} \left( \frac{\delta q_T^{wp}}{\delta x} \right)_{+x} = -\frac{1}{6} u^{wp} \left\langle \Lambda_{wp} \right\rangle \frac{\partial q_T^{wp}}{\partial T} \left\langle \frac{\delta T}{\delta x} \right\rangle_{+x}$

is generated along the "+$x$" direction, where $D_+^{wp} = \frac{1}{6} \left\langle u_{wp} \right\rangle \left\langle \Lambda_{wp} \right\rangle$ is the wave packets diffusion coefficient along "+$x$". Because in an isothermal medium the number of scatterings along any direction is balanced by that in the opposite one, to get a null diffusion over time, equations similar to Eqs.(5) and (6) must be set up for the opposite direction "–$x$"; the algebraic summation of the "+$x$" and "–$x$" equations gives the usual diffusion coefficient $D^{wp}$

7. $D^{wp} = \frac{\left\langle \Lambda_{wp} \right\rangle \left\langle u_{wp} \right\rangle}{3} = \frac{\left\langle \Lambda_{wp} \right\rangle^2 / \left\langle \tau_{wp} \right\rangle}{3} = \frac{\left\langle \Lambda_{wp} \right\rangle^2 \left\langle \nu_{wp} \right\rangle}{3}$

for the wave-packets. On the contrary, when an external thermodynamic (generalized) force is applied to the system, as for instance a temperature or a velocity gradient, as it is the case for the viscosity, a flux (in the opposite direction) is generated, and therefore an increase of "wave-packet ↔ *liquid particle*" interactions in the same direction [41,44].

Let us now take in consideration the case of a liquid contained in between two walls, with one of the two walls moving with respect to the other with a relative velocity $d\upsilon_x$, in such a way that a linear velocity gradient $\frac{d\upsilon_x}{dy}$ is applied to the liquid along the $y$-direction in stationary conditions, as in Figure 1d). For each of the layers in which the liquid may be ideally subdivided, there will be an equal number per second of elementary interactions of type a) and b) of Figure 2 along the $x$-direction, so that their effect is null in the average along that direction. However, the average number per second of elementary interactions is not the same for all the layers, due to the presence of $\frac{d\upsilon_x}{dy}$. Let's consider, for instance, a layer '1' moving at speed $\upsilon_1$ and an adjacent layer '2' moving at speed $\upsilon_2$, with $\upsilon_2 > \upsilon_1$, as shown in Figure 3. Taking in mind the Eq.(1), the viscosity is the friction acting between the two layers. This force, as is well known, works like a shear stress, as the fast layer pushes the slow one. In the DML, however, such friction is mediated by the presence of the wave-packets, responsible for the momentum (and energy) transport pertaining to the solid-like collective DoF. In Figure 3, a *liquid particle* A belonging to the Layer #2, moves at speed $\upsilon_2$ through the liquid in the $x$-direction, and a *liquid particle* B belonging to the Layer #1, moves at speed $\upsilon_1$ always in the $x$-direction. The *liquid particle* A is in an excited state and delivers an energetic wave-packet following the process b) of Figure 2; this crosses the border layer and hits the *liquid particle* B. As consequence of the emission, the *liquid particle* A goes in a de-excited state and slows down, due to the recoil. The opposite fate occurs to the *liquid particle* B: it acquires momentum and energy as consequence of the interaction with the wave-packet, as in an event a) of Figure 2. The acquired momentum and energy increase its velocity and the internal energy, exciting the collective DoF. The kinetic energy will be dissipated by means of collisions with the other particles of Layer #1, while the energy acquired by the internal DoF will be released into the thermal reservoir of Layer #1 (the reader interested to the relevance and to the consequences of the tunnel effect characterizing the *liquid-particle* $\leftrightarrow$ wave packet interaction, may refer to [41,43]). Through this physical mechanism, much more complicate to describe than to understand, momentum and energy are transferred from layer-to-layer by means of the wave-packets, from those with larger momentum density down to those with lesser momentum density.

Let us translate now the above idea into a formal building using the laws of physics. We begin by noting that Eq. (1), relating the shear force per unit area, $f_{x,y}$, between two adjacent liquid layers to the generic external velocity gradient $\frac{d\upsilon_x}{dy}$, may be interpreted also in another fashion. In the neighbourhood of the moving surface at $y = 0$ of Figure 1, the fluid acquires a certain amount of $x$-momentum. This fluid, in turn, imparts shear momentum to the adjacent liquid layer by means of the wave-packets, causing it to move even in the $x$-direction. Hence, $x$-momentum is being transmitted through the fluid in the $y$-direction; consequently, $f_{x,y}$ may also be interpreted as the flux of $x$-momentum in the $y$-direction. This interpretation is consistent with the molecular picture of momentum transport and the kinetic theories of gases and liquids, and does not need of intricate mechanisms to allow momentum being transported through the liquid. It is also in harmony with the analogous treatment given for heat and mass transport in the DML, in particular for that used to interpret the mechano-thermal effect in liquids under shear [44]. This idea copes with the common thought that momentum flows from a region of high velocity to a region of low velocity, just as the way heat flows from a region of high temperature to a region of low temperature. The velocity gradient can therefore be thought of as the "driving force" for momentum transport, and the momentum flux as the "generalized flux" in the usual meaning of the NET. Let us then build the equivalent of Eq.(1) for the momentum flux carried by the lattice particles. The first observation is that, since the motion of the wave-packets driven by $\frac{d\upsilon_x}{dy}$, crossing the boundary between two adjacent layers, is an ordinary diffusive process for the wave-packets, it is described by the diffusion coefficient $D^{wp}$ introduced in Eq.(7). The product of $D^{wp}$ by the density $\rho_l^{lp}$, defines the dynamic viscosity $\eta_l^{wp}$, i.e. the momentum per unit of surface:

8. $\quad \eta_l^{wp} = \rho_l^{lp} \frac{\langle \Lambda_{wp} \rangle^2 \langle v_{wp} \rangle}{3}$

It should not be surprising that eq. (8) is gas-kinetic like. Here we are in fact evaluating only the contribution due to the interaction between the two statistical populations of pseudo-particles that make up the liquid in the DML. Therefore, we are neglecting either the inter-molecular interactions, or phenomena involving multiple *lattice particle* interactions. We will return on this aspect in the Discussion. Because we are considering here only the momentum transfer mediated by the wave-packets, the density $\rho_l^{lp}$ accounts only for how many *liquid particles* are present per unit of volume, $N_{lp}$, and for their average mass, $m_{lp}$,

9. $\quad \rho_l^{lp} = m_{lp} N_{lp}$.

Of course, it holds $\rho_l^{lp} < \rho_l$. As we have extensively discussed in previous papers, especially in [41], wave-packets ensure the momentum and energy transport in the DML only through their interactions with the *liquid particles*. It is then instinctive and straightaway to replace Eq.(1) in DML with

10. $\quad J_p^{wp}(y) = -\rho_l^{lp} D^{wp} \frac{d\upsilon_x}{dy} = -m_{lp} N_{lp} \frac{\langle \Lambda_{wp} \rangle^2 \langle v_{wp} \rangle}{3} \frac{d\upsilon_x}{dy} = -\eta_l^{wp} \frac{d\upsilon_x}{dy}$

for the momentum flux (formally coincident with the energy density) carried by wave-packets in the $y$-direction, driven by the velocity gradient $\frac{d\upsilon_x}{dy}$. Equation (10) is the typical equation that in NET relates the momentum flux with the gradient of velocity; it has a structure similar to any other diffusion equation in NET, such as the Fourier equation for the heat flow, the Ohm's law for the electric charges flow or the Stokes equation for mass flow, for instance. Equation (8) defines the viscosity $\eta_l^{wp}$ following the usual meaning of the momentum transported per unit of surface by the wave-packets. The linearity of Eq. (10) ensures that we are dealing with Newtonian fluids. In what follows we shall sometimes refer to Newton's law in terms of forces (which emphasizes the mechanical nature of the subject) and sometimes in terms of momentum transport (which emphasizes the analogies with heat and mass transport). This dual viewpoint should proves helpful in the physical interpretation. The proposed model is applicable also to the drag viscosity between a solid surface and a liquid adjacent to it [100,121]. We will return on this point in the Discussion.

## 3.2 The $T$-dependence of $\eta_l^{wp}$

To complete the tasks of the manuscript it lefts to show that $\eta_l^{wp}$ has an Arrhenius-like behaviour vs temperature. The Arrhenius equation, named after the Swedish chemist who introduced it, later reworked by Eyring as the Rate Reaction Theory, describes the kinetics of linear chemical reactions vs temperature, $K = K_0 e^{-\Delta e/K_B T}$, where $K$ is the kinetic constant that characterizes the chemical reactions. The rate of a chemical reaction measures the quantity of matter consumed (reactants) or produced (products) over time because of a chemical reaction. The speed of a reaction generally depends on the concentration of the reactants, the temperature and the activation energy, i.e. the energy threshold above which the reaction takes place. Applying this concept to the problem of viscosity, $\eta_l^{wp}$ takes on the role of the kinetic constant, being a function of the concentration, and size, of the icebergs and wave-packets. To accomplish the last task, firstly we recall the reader's attention on the fact that $\eta_l^{wp}$ depends on three quantities, namely $\rho_l^{lp}$, $\langle v_{wp} \rangle$ and $\langle \Lambda_{wp} \rangle$. For $\eta_l^{wp}$ to exhibit an Arrhenius-like trend, the above quantities would have the same trend, or at least a decreasing trend vs $T$. Let's start then with $\rho_l^{lp}$. Although ubiquitous in liquids, the number and size of the *liquid particles* (as well as of the *lattice particles* and wave-packets) decrease with temperature, being them related to the parameter $m$, with $0 \leq m \leq 1$ and $\frac{dm}{dT} < 0$ [41-42]. A dedicated paper on the topic of the statistical evaluation of the such parameter is in preparation [Peluso, F., et al.], however it is relevant for the time being to quote the results achieved by Ghandili et al. [122] on a similar topic. They simulate a liquid in a similar way as the DML, i.e. as a system arranged by means of molecular clusters and free particles. The molecules within the clusters may experience only quasi-harmonic vibrations around their equilibrium positions. The authors introduce two parameters to describe the characteristics of the system as function of the temperature and/or the number of DoF excited. One is the concept of thermodynamic dimension, $D_T$, corresponding to the average number of intermolecular interactions per *liquid particle* as function of the absolute temperature $T$. They show that $D_T$ turns out to be a measure of how many molecules are bound in the lattice of a single *liquid particle*. The authors show that their approach distinguishes free particles from bound particles. The other parameter is $\Theta_{max}$, for the liquid temperature corresponding to the limit at which all DoF begin to be excited. It is clear that $\Theta_{max}$ is linked to the parameter $m$ introduced in the DML [41]. At temperatures below or around $\Theta_{max}$, the system is characterized by $D_T = 3$, while this parameter goes to zero as the temperature increases (analogously to $m$). The authors propose a general expression for the vibrational DoF contribution to the isochoric specific heat (in a similar way as we have done in [41-42]), and show that in a wide range of temperature and pressure, the calculated results agree very well with experimental data for isochoric heat capacities in dense regions in the considered fluids. In particular, by introducing a new condition ($D_T = 1/2$) to plot the Frenkel line, they predict solid-like features around the critical point. The above argument supports our hypothesis that an Arrhenius-like functional dependence for $\rho_l^{lp}$ with respect to temperature may be reasonably supposed, or at least a decreasing trend vs $T$, so that the momentum flux $J_p^{wp}(y)$ exchanged between *liquid particles* by means of the collisions with the wave-packets, decreases with temperature because the same do the number of icebergs and of the wave-packets. For the moment, we propose in the DML the following linear "reaction rate" equation

11. $\dfrac{d\rho_l^{lp}}{dT} = -\dfrac{\rho_l^{lp}}{T}\dfrac{\Delta E}{K_B T}$,

for the density of wave-packets vs $T$, where $\Delta E$ is, as usual, the activation energy of the interaction. By means of variable separation technique, it is trivial to integrate Eq.(11) and get the Arrhenius-like equation for $\rho_l^{lp}$.

Cunsolo et al. have shown, on the basis of experimental results and Molecular Dynamic simulations, that in water [82-83,85,94,96] and other liquids [86,97], the relaxation time $\langle \tau_{wp} \rangle$ shows an Arrhenius plot with positive slope vs $T$. Because $\langle \tau_{wp} \rangle = 1/\langle \nu_{wp} \rangle$, $\langle \nu_{wp} \rangle$ shall exhibit the expected Arrhenius behaviour, with a negative slope. This behaviour is easily understandable as due to two circumstances working at the same time at mesoscopic scale, i.e. the decreasing density of the *liquid particles*, and the decreasing number of collective DoF, i.e. the *lattice particles*, due to the condition $dm/dT < 1$, and consequently of the wave-packets emitted by the *liquid particles*. Definitively, we may assume that also for $\langle \nu_{wp} \rangle$ it holds:

12. $\dfrac{d\langle \nu_{wp} \rangle}{dT} = -\dfrac{\langle \nu_{wp} \rangle}{T}\dfrac{\Delta E}{K_B T}$

where we used for simplicity the same symbol for the activation energy as in Eq.(11).

As it concerns $\langle \Lambda_{wp} \rangle$, it is necessary to make a wider reasoning, by starting to recall the dynamics occurring in a liquid following the DML. A wave-packet travels at the speed $\langle u_{wp} \rangle = \dfrac{\langle \Lambda_{wp} \rangle}{\langle \tau_{wp} \rangle}$ until hitting a *liquid particle*, as in Figure 2. A fraction of the energy lost by the wave packet is commuted into kinetic energy of the cluster, part excites its internal quantized DoF (harmonic contribution). The perturbation inside the solid-like structure propagates by means of (quasi) harmonic waves, travelling at speed $\langle u^0 \rangle = \frac{\langle \Lambda_0 \rangle}{\langle \tau_0 \rangle}$. The DML provides currently the OoM of several parameters, evaluated in a previous paper [41]; among them, $\langle \Lambda_0 \rangle$ and $\langle \tau_0 \rangle$. Both these quantities are related to the dynamics involving the *internal* collective DoF of the *liquid particles*. In the DML the wave-packets represent the long-ranged medium by means of which the icebergs communicate, exchanging energy and momentum each other and with the phonons themselves. Wave-packets travel through the amorphous phase of the liquid; their anharmonic character allows momentum be transferred after the collision, and their lifetime is finite as well as their mean free path. Therefore, although we do not have yet a direct way to evaluate either $\langle \Lambda_{wp} \rangle$ or its $T$–dependence, on the basis of the above reasoning based on the decreasing trend of phononic contribution to the global heat content of a liquid, it is reasonable to expect that even for it a decreasing trend vs temperature is expected. $\langle \Lambda_{wp} \rangle$ is in fact strongly related to the correlation length in liquids (in the DML the correlation length is linked to both $\langle \Lambda_{wp} \rangle$ and $\langle \Lambda_0 \rangle$). In [41, and references therein] we have extensively discussed and evidenced how the measurement of the correlation length in liquids decreases with increasing the temperature, i.e. from the solid-like value to the liquid typical one, i.e. from $\approx 3200\ m/s$ to $\approx 1500\ m/s$ for the case of water. We have shown above that the two other quantities entering in the definition of $\eta_l^{wp}$ show an Arrhenius-like dependence; therefore we may deduce an Arrhenius-like dependence also for $\eta_l^{wp}$ (or, however, at least a decreasing trend vs $T$).

## 4. Comparison of the Theoretical Results with the Experimental Data and Recent Theoretical/Numerical Modelling.

What remains to do now is supporting the model by comparing the theoretical results with the experimental data. Before proceeding with the topic, it is worth pointing out that $J_p^{wp}(y)$ could be a large or a small part of the total momentum flux, $J_p(y)$, as well as $\eta_l^{wp}$ a large or a small fraction of the total viscosity, $\eta_l$; however, this is a question that can only be answered with dedicated experiments, at least until a suitable model for the parameter $m$ is available (the same warning was given in the previous papers where we have obtained the expressions for the specific heat or the thermal conductivity of liquids due to the wave-packets contribution [41-43]). Let's then start with the numerical evaluation of the OoM of $\eta_l^{wp}$ based on Eq.(8). Due to the lacking of theoretical values for $\langle\Lambda_{wp}\rangle$, $\langle\tau_{wp}\rangle$ and $\rho_l^{lp}$, we shall make use of the values deduced for $\langle\Lambda_0\rangle$ and $\langle\tau_0\rangle$ for the case of water, reported in [41,44], as well as we use the standard density of water instead of $\rho_l^{lp}$. Table 1 summarizes the results, where we have used the minimum and maximum values for the mean free path and the lifetime of *lattice particles*. The reader can see that, notwithstanding all the intrinsic limitations in the choices either of the harmonic quantities in place of the anharmonic ones, or of their values, the values obtained for the dynamic viscosity of water are within the range of one OoM [68], taking even in mind that the value of $\rho_l$ was used instead of $\rho_l^{lp}$. We are aware and recognize that at present we cannot go further than an OoM evaluation of $\eta_l^{wp}$, and that only suitably designed experiments may provide accurate evaluations of the several quantities involved in its expression. However, the reasonable agreement between the theoretical predicted values and those available from experiments for $\eta_l$, allows us to state that a significant fraction of the viscosity in liquids, i.e. of the momentum exchanged between *liquid particles*, is due to the physical mechanism described in this manuscript. One could also attempt a speculative use of the Eq.(8) exploiting it in a reverse way, i.e. by deducing the maximum values for $\langle\Lambda_{wp}\rangle$, $\langle\tau_{wp}\rangle$ and $\rho_l^{lp}$ from the experimental values of $\eta_l$. On the other side, we can't help but remember that it is usually accepted as a good result in the theories of viscosity of liquids when the theoretical values are in agreement with the experimental data within one OoM and with the main trends of the dependence on molecular properties and external parameters (see [68], chapter 2.2 for an in-depth discussion on this topic). All theories developed so far have only been able to reproduce experimental data without predicting any essential new facts or correlations valid in a broader range. For some liquids of simple structure under definite conditions, Andrade [47-50] and Kirkwood [51-53] deduced numerical values for the viscosity by using only molecular constants known from other phenomena; nevertheless, these results could not be generalized without introducing empirical parameters. Therefore, they cannot serve as a basis for a general theory of the viscosity of liquids [68,123 -125].

An even more robust test bench of the model of viscosity in the DML is represented by the physical interpretation of the phenomenology discovered in liquids under shear strain by Noirez and co-workers [44,1-9]. The phenomenon consists in an "unexpected" mechano-thermal effect that manifests in isothermal liquids in shear motion [1-9]. It was detected in two types of similar experiments, whose difference is only the motion law of a movable disk, oscillating or following a Heaviside function. The effect consists, in both cases, in the occurrence in a liquid under shear of a temperature gradient, opposite to the shear gradient, due to the momentum transferred to the liquid by a moving plate. This mechano-thermal effect reveals a coupling between the mechanical force, induced to the liquid by the shear strain, and the thermal gradient generated in the liquid dragged by the shear motion. The universal law $G' \approx L^{-3}$ for the low-frequency shear modulus $G'$ was confirmed in confined liquids [5,6], however, the first-ever theoretical explanation of such phenomenology at mesoscopic level was very recently proposed by us in the DML frame [44]. Here we generalize this theoretical interpretation of the phenomenology within the modelling of the viscosity provided in the present manuscript.

Based on the classical fluid dynamics and with reference to Figure 5 in [44], one should expect that part of the work of the external force needed to overcome the internal liquid friction, i.e. the viscosity, is transformed into thermal energy. As consequence, the momentum transfer from the moving plate should induce a general heating of the liquid, in particular of the liquid layers facing the moving plate. A first problem already arises because in reality one would not even expect a faint heating of the liquid layer. Indeed, at such low speeds, the stored energy cannot generate viscous heating according to the empirical evaluation based on the Nahme (or Brinkman) number, since for viscous liquids much higher speeds are needed, close to that of sound.

In the DML framework, the phenomenon is framed by a different approach. Let us then return for a while to Figure 3, which represents what happens inside a liquid layer under shear. Because of the duality of liquids, the energy is transferred through the *liquid particle* ↔ wave packet collision, from the fast layer to the slow one. This induces a heating of the slow one, in particular of that part of each layer close to the slower one, and not of the layer closer to the mobile plate. In the DML, in fact, one expects the onset of a thermal gradient, oriented in the opposite direction to the velocity gradient, i.e. from the faster layers toward the slower ones, as shown in Figure 4, due to the current of wave-packets, carrying momentum and energy. They are pushed by the external shear force from the fast to the slow layers where they accumulate, in particular to the far side of the slow layer with respect to the fast one. We can't help but notice that this crossed effect is exactly the phenomenology highlighted by Noirez in isothermal pure liquids under shear motion [1-9], that was identified as "unexpected". While, from the NET point of view, this phenomenon can be assumed as the generalized response of the system to the external disturbance consisting in the momentum gradient, the DML provides the first-ever physical interpretation of the phenomenon discovered by Noirez, introduced in a previous paper [44], and quantifiable, as we will show now. The external generalized force in this case is the velocity gradient $\frac{d\upsilon_x}{dy}$, that generates the momentum flux $J_p^{wp}(y)$ given by Eq.(10). The system reacts with a faint thermal gradient in the opposite direction, as shown in Figure 4. Following the above reasoning, it is possible to get the expression for the temperature difference generated by the wave-packet' current given by Eq.(10). Other than the momentum flux, $J_p^{wp}(y)$ represents also the density of thermal energy due to the wave-packet' current. Therefore, dividing $J_p^{wp}(y)$ by the heat capacity, one gets the searched $\Delta T$:

13. $$\Delta T = \frac{\overline{J_p^{wp}(y)}}{(\rho C)_l} = \frac{W_{wp}}{(\rho C)_l} = \eta_l^{wp} \frac{\rho_l^{lp}}{(\rho C)_l} \overline{\frac{d\upsilon_x}{dy}}$$

where we have used the "total" heat capacity of the liquid because the heating affects the whole liquid, as well as the vector module '‾‾‾' to avoid any sign misunderstanding. The physical process described in the previous section and in the Figure 4 continues until the velocity gradient $\frac{d\upsilon_x}{dy}$ is working, the viscosity representing the manifestation of the tendency of any system to reach the equilibrium. Very interestingly, in [44] we calculated the amount of macroscopic momentum flux injected in the liquid due to the plate rotation, obtaining an expression analogous to eq.(13), that is instead obtained by through the definition of viscosity introduced in the present manuscript. The equality of the two expressions is a confirmation of the rightness of the approaches. The numerical estimation of $\Delta T$ in eq.(13) can be calculated by means of the data contained in [2], as we did in [44], so that one gets $\Delta T \approx 0.06K$, of the correct order of magnitude as measured and reported in [2]. This phenomenology, at the best of the author' knowledge, has never been evidenced in liquids (but also never looked for!, except incidentally from the Noirez' group), we may conclude that even this aspect of the physical mechanism proposed for the viscosity in flowing liquids due to their duality, is explained by the DML. The onset of a temperature gradient opposite to the velocity gradient in the liquid is therefore not unexpected at all, but rather a crossed effect, typical in NET, absolutely expected in systems under the influence of an external force field.

In conclusion, in the DML, the viscosity is nothing else than a process that tends to balance the *liquid particles* and wave-packets populations with different momentum contents.

Ghandili has recently proposed an ingenious idea to modelling the viscosity in liquids following a thermodynamic approach [126]. The fluid is considered as a fractal lattice, in which temporary molecular solid-like clusters, dubbed *t-clusters*, dive. This is once again an approach very similarly to that on which DML is based. The author recalls the importance of the relaxation time as pivotal for the modelling of the behaviour of liquids on mesoscopic scale: in liquids, molecules either vibrate around their equilibrium positions, similarly to solids, or jump from one place to the nearest neighbour position, similarly to gases. The vibratory property vanishes as temperature rises, as well as the number and extension of the solid-like temporary clusters. Ghandili reused the intriguing concept of thermodynamic dimension $D_T$ previously introduced to model the heat capacity [122]. Similarly to the parameter $m$ in the DML, the author introduces two distinct types of DoF, those pertaining to the free particles of the liquid, and those pertaining to the molecules bound in the clusters. The modelling of the contribution to the viscosity due to the solid-like clusters goes through the calculation of the force and related energy to strain and move them. This approach, in turn, allows splitting the viscosity in liquids as the sum of two components, one accounting for the *t-clusters* displacement, and therefore proportional to their kinetic energy variation, alike in a gas, the second accounting for their deformation, and therefore proportional to their internal potential energy variation, alike in a solid (the reader may recognize as many points are shared with the DML [41-44], in particular, it is worth noticing that the fluctuations of internal potential energy is associated to a fraction of the viscosity). The author shows that numerical elaboration of the equations allows getting a very good agreement in liquid $N_2$ between the viscosity data and the theoretical results, and therefore even an Arrhenius trend vs temperature.

The results obtained by Ghandili represent a strong and reliable support for the DML, too.

## 5. Discussion.

Despite the fact that the physical origin and the law of the viscosity of gases have been known for a long time, our theoretical knowledge on the viscosity of liquids is much more incomplete because of the more complicated and lesser-known structure of liquids if compared with that of gases. Besides the fact that liquids "do not have small parameters", as is well known the physicists have long been divided into two factions: the supporters of liquids considered as condensed gases, and the supporters of liquids considered as molten solids [127]. If today a definite and general view of the mesoscopic organization of liquids is provided by the experimental data, Eckart was the first to explicitly assert that "the distinction between liquids and solids is quantitative and not qualitative" in his papers on the anelastic fluid [128-129], the border line being the relaxation time $\tau_F$, or the frequency $\nu_F$. Eckart obtained in his papers many other interesting results, for instance that the dependence on frequency makes the propagation velocity of isentropic longitudinal waves a complex number. The problem of the relaxation times occurring in the dissipative processes was faced by neglecting all the possible causes of dissipation except the collisions (as in the DML); this approach gave origin to the viscosity. Trachenko [130] raised an interesting connection between such characteristic of the liquids and the property of phonons travelling through them: unlike in solids, the phonons in liquids have the relevant property that their phase space is not fixed but variable; in particular, the phase space reduces with temperature. This could explain why, for instance, the specific heat in liquids decreases with temperature. In the DML such behaviour is due to the trend of the parameter $m$ vs $T$, $dm/dT < 0$. The above reasoning confirms the close relation between $m$ and the VDOS [41,79-80,118].

Various models were proposed in the past century for the internal friction of liquids and liquid solutions, but some of them give no more than a qualitative interpretation of the empirical equations and of the several factors in them, while others, partly on the basis of quantum mechanics, try to correlate the phenomena of viscous flow with the properties of the atoms or molecules by means of an extensive mathematical treatment. Furthermore, what jumps to the reader' attention is the fact that the several models are based on as much different physical mechanisms responsible for the internal friction of liquids. Therefore, it is not surprising if none of them can be regarded as a satisfactory theory [68], so far only the theory of the viscosity of liquids of the simplest structures, consisting of monatomic molecules with spherically symmetrical fields of very short range, i.e. the noble gases, has been successfully elaborated [51-53,123-125]. To them we may add the recent works by Zaccone [79-80] and Ghandili [126], both based on a dual modelling of liquids on mesoscopic scale.

Notwithstanding the Maxwell model of viscosity of gases cannot be tailored for the case of liquids, one should not be surprised that eq. (8) for $\eta_l^{wp}$ is a gas-kinetic like expression, for several reasons. Firstly, because the goal is to provide a physical model only for that part of the viscosity of liquids related to the presence of inelastic wave-packets (dual liquids), $\eta_l^{wp}$, here we deal with only the consequences of the momentum transferred through the interactions between *liquid particles* and wave-packets. In the present case, the wave-packets emerging from every single *liquid particle*, play the role of carriers and exchangers of momentum between the various liquid layers in relative motion, or better, under a shear motion. Therefore, on one side we neglect the inter-molecular interactions, the phenomena involving multiple *lattice particle* interactions (Umklapp), as well as multiple wave-packets ↔ *liquid particle* interactions. On the other side, eq. (8) is daughter of the hypothesis that the two populations interact only "if" and "when" they come into contact, interactions at distance being not accounted for [41]. The above arguments clarify why we adopted a gas kinetic modelling for $\eta_l^{wp}$. However, it becomes relevant to have reasonable arguments on which to establish the extension of the wave packets. Although this is not a trivial task, we have already tentatively schematized it through the definition of $\langle \Lambda_{wp} \rangle = n \lambda_{wp}$ [see section 3.4 in 41]. Whether the contribution of $\eta_l^{wp}$ is a small or large part of the "whole", can only be established through specifically designed experiments (see Conclusions), theoretical arguments or numerical tools that can resolve this aspect being still currently "work in progress". The model illustrated in the present manuscript, as well as the numerical evaluation provided by Ghandili [126] indicate however that the way taken is the correct one. Of course, an accurate evaluation of the parameter $m$, possibly even in connection with the VDOS and/or the thermodynamic dimension $D_T$, can certainly give clear indications on the relative importance of this contribution [Peluso et al, in preparation]. If experiments aimed at determining the contribution of $\eta_l^{wp}$ with respect to the total viscosity were to be carried out in the future, the experimental results would also provide an intrinsic evaluation of $m$.

A second argument supporting eq. (8) is that, whatever the specific dependence of all the quantities involved, we have shown that for each of them the contribution to $\eta_l^{wp}$ decreases with $T$. In the DML, in fact, although the wave-packets are dealt with as a gas, the momentum exchanged between the wave-packet' population and the *liquid particles* decreases with temperature because the same do their statistical population; therefore, their contribution becomes less and less relevant in the global computation of the viscosity. In support of these arguments it turns out relevant the fact that the density of wave-packets $N_{wp}$ may be estimated in water at room conditions as $N_{wp} \approx m \cdot 4 \cdot 10^{22} \, cm^{-3}$ [41]; because $\frac{dm}{dT} < 0$, such density decreases with $T$ (of course, a similar dependence upon $m$ is expected in other liquids, only the numerical factor can change). Once again, the proposed model for $\eta_l^{wp}$ shall be assumed as additional to those currently accepted for $\eta_l$, in particular for liquids at low temperature and/or high pressure, where the specific contribution due to the transport of momentum and energy by wave-packets is more relevant.

It is worth noticing that any "classical" model of viscosity in liquids does not allow to get an Arrhenius behaviour directly from the theory, because there is no physical mechanism that foresees the interacting molecules to decrease their number with temperature (with the only exception of the thermodynamic approach by Ghandili [126]). The Eyring model, for instance, loans from chemistry the concept of holes not only to allow molecules to move, but also to have a way to reproduce the Arrhenius trend. The DML does not need of any additional mechanism because the momentum transport between *liquid particles* with different momentum is mediated by the elastic wave-packets, whose concentration depends on how many collective DoF are available at a given temperature, given by the parameter $m$, that decreases with $T$. It would not be a surprise if further investigation about $m$, $N_{wp}$ or $N_{lp}$ revealed an Arrhenius-type dependence on $T$ for these quantities as well.

Although we have considered in this manuscript only the momentum carried by the wave-packets travelling through the liquid at speed $\langle u_{wp} \rangle = \frac{\langle \Lambda_{wp} \rangle}{\langle \tau_{wp} \rangle}$, and exchanged with the *liquid particles* belonging to adjacent layers, we have used in the Eq.(10) for the momentum flux $J_p^{wp}(y)$ the velocity gradient of the (relative) shear velocity $\frac{d\upsilon_x}{dy}$ instead of a gradient of $\langle u_{wp} \rangle$. The reason is that $\langle u_{wp} \rangle$, by definition of the average, is the same throughout the liquid, because we are neglecting any external effect except the shear velocity (isothermal system) so that it cancels in calculating the relative speed, only second order effects due to a faint ($\upsilon_x \ll \langle u_{wp} \rangle$) Doppler shift remaining (the Doppler shift becomes relevant when energetic neutrons or photons are injected in a liquid, giving rise to the well-known Brillouin triplet in the homologous scattering). The only velocity gradient that matters is therefore $\frac{d\upsilon_x}{dy}$.

We want to attract the reader' attention also on the fact that the mean free path of a wave-packet, $\langle \Lambda_{wp} \rangle$, should not be confused with the average intermolecular distance, $l$, also depending upon the temperature, although with a reverse trend. Notwithstanding these concepts were extensively discussed by many authors [82-83,85-86,92] and by us [41], we report here the driving points to give the reader a correct general vision of the statement. When dealing with crystalline systems, the investigation of the translational invariance is reduced to the small number of atoms constituting the crystal unit cell. This simplification allowed to considerably simplify classical and quantum modelling of crystalline solids, from both the theoretical and numerical points of view. However, when the crystalline order is left, such as in disordered systems as liquids, relevant differences occur. Disordered systems are in fact characterized by two conceptually different length scales, namely the inter-particle distance, $l$, and the correlation length, $\lambda$, this last characterizing the so-called topological disorder. The understanding of the particle dynamics in disordered (or ordered over short distances) systems is complicated not only by the absence of translational invariance, but also by the presence of additional DoF, that characterize the diffusion, the relaxation phenomena as well as the tunnelling (we have generalized these additional DoF in the DML by introducing $m$). These processes necessitate the introduction also of an additional time scale, $\tau$, on the base of which the several relaxation times are compared among them. In liquids, two limiting cases may be envisaged, namely those of the excitations with characteristic space and time scales which are either very long or very short compared with $\lambda$ and $\tau$. They are, respectively, the hydrodynamic and the single-particle limits. In contrast, in the intermediate, so-called mesoscopic scale, an exhaustive modelling is still missing, although recent theoretical developments and experimental investigations are progressively filling such gap (and the DML is hopefully contributing to this effort).

We want now to linger on the ratio between the thermal conductivity $K_l$ and the specific heat $C_l$. This ratio, when referred to the "total" quantities, represents the dynamic viscosity, $\eta_l = \frac{K_l}{C_l}$. Despite the fact that it was early introduced for gases [131, chapter 5, figure 5.1 and Eq. 5.76)], it was proven valid also in the case of liquids [132, Chapter 12.6, eq. 12.20b]. We want submit to the reader' attention some considerations on the analogous expression but referred to the dual system,

14. $\eta_l^{wp} = \frac{K_l^{wp}}{C_l^{wp}}$,

in particular to the gas of wave-packets. Because we have already calculated the expressions for $K_l^{wp}$ and $C_l^{wp}$ in previous papers [42-43],

15. $K_l^{wp} = \frac{1}{3}\langle u_{wp}\rangle\langle \Lambda_{wp}\rangle\frac{\partial}{\partial T}\left[\mathcal{N}^{wp}\langle\varepsilon^{wp}\rangle\right] = \frac{1}{3}\langle u_{wp}\rangle\langle \Lambda_{wp}\rangle m\rho C_V\left[\frac{m^*}{m^2}\frac{dm}{dT}T+1\right] = \frac{1}{3}\langle u_{wp}\rangle\langle \Lambda_{wp}\rangle\rho C_l^{wp}$

16. $\begin{cases}\rho C_l^{wp} = \frac{\partial q_T^{wp}}{\partial T} = \frac{\partial}{\partial T}(mq_T) = q_T\frac{dm}{dT} + m\frac{\partial q_T}{\partial T} = q_T\frac{dm}{dT} + m\rho C_1 = \\ = m\rho C_1\left[\frac{q_T}{m\rho C_1}\frac{dm}{dT}+1\right] = m\rho C_V\left[\frac{m^*}{m^2}\frac{dm}{dT}T+1\right]\end{cases}$

it is easy to verify that, by inserting them into Eq.(14) and remembering that $\langle u_{wp}\rangle = \langle \Lambda_{wp}\rangle\langle \nu_{wp}\rangle$, Eq.(8) is readily obtained. Furthermore, we have highlighted [43] that $\eta_l$ appears in the propagation equation for the thermal energy carried by wave-packets in the DML,

17. $\frac{\partial^2 u}{\partial z^2} = \frac{\rho}{\eta_l}\frac{\partial u}{\partial t} + \tau\frac{\rho}{\eta_l}\frac{\partial^2 u}{\partial t^2}$

where $u$ is the fluid velocity component normal to $z$. We advanced the speculative hypothesis that $\eta_l^{wp}$ could represent the capability of wave-packets to diffuse momentum, in the same way as $\eta_l$ does for the pure molecular fluid [43, Eqs.(24) and (15) therein]. In light of what we have supposed to be the mechanism by which momentum is spread through wave-packets (Figure 3), it is possible to provide another physical interpretation of the ratio of Eq.(14). The wave-packet is emitted by a fast *liquid particle*, as such it may be considered as a "quantum" of thermal conductivity; after the impact with a slow *liquid particle*, it is absorbed by it as a "quantum" of specific heat. Therefore, $\eta_l^{wp}$ could be considered as the ratio between the heat emitted by diffusion by a *liquid particle* and that absorbed, and as such it may be considered the analogous of (the reciprocal of) the Prandtl number for the current of wave-packets. In the same way, we may intuitively identify the kinematic viscosity $\xi_l^{wp}$ for the momentum diffusivity due to the gas of wave-packet' component with their diffusivity coefficient given by Eq.(7), $\xi_l^{wp} \equiv D^{wp}$.

It is mandatory to spend some words on the nature of the viscosity activation energy in the dual liquid framework. The physical mechanism of viscosity in the DML should be compiled with the wider reasoning about the nature of the wave-packet ↔ *liquid particle* collision [41]. Apart from this, for the collision to have a dynamical effect, the wave-packet emitted from the fast *liquid particle* must have energy and momentum large enough to a) accelerate the slow *liquid particle*, and/or b) excite its internal collective DoF. The two effects are of course independent and, in principle, not mutually exclusive. The evaluation of the two energy levels is not exactly feasible yet, because it is a task that necessitates of either dedicated theoretical or accurate numerical modelling (some elements have already been provided in [41] for the evaluation of the OoM of the relaxation times involved in the interaction). This notwithstanding, some insights can already be advanced.

Among the two energy reservoirs, the kinetic and that affecting the internal collective DoF, the former, that has a classical nature, is probably the less effective in the balance of the viscosity, especially at low temperature and high pressure, because of the limited capability of molecules, and of *liquid particles*, to move over large distances as consequence of the interaction. Therefore, the second, that has a quantum nature, is probably the more effective among the two in the balance of the viscosity. A similar conclusion was get also by Ghandili [126] although based on a different argument (we should also take in consideration that the viscosity of gases is based on a pure kinetic effect for the propagation of momentum, and it does not provide an Arrhenius trend). This argument, at moment purely speculative, would provide a key role to the (classical) tunnel effect characterizing the wave-packet $\leftrightarrow$ *liquid particle* collision in the DML (see Figure in 4 [41]), as well as to the collective internal DoF, excited by the *lattice particles*. For the above reasons, it is reasonable to assume that the activation energy has mostly a quantum nature and is proportional to the parameter $m$. The quantum nature should therefore prevail on the classical one as the temperature decreases. This conclusion allows us to highlight on the role that fundamental physical constants have on the fact that viscosity appears to have a universal lower limit, as pointed out by Trachenko et al. [11,133]. They deduce that a combination of some fundamental physical constants of the nature may both give a minimal kinematic viscosity and an elementary viscosity, both depending, among others, upon $h$ and $K_B$. As the authors highlight in their papers, the presence of these two constants is the fingerprint of the quantum nature of the viscosity. The expression of the dynamic viscosity in the DML is given by eq.(8) and related eqs.(11) and (12), where we may recognize the presence of both $K_B$ and $h$. The first is explicitly reported in the denominator at the exponential, while the second is hidden at least into the activation energy, that is related to the energy exchanged between the wave-packets and the *liquid particles*; such energy shall be surely expressed as a function of the energy exchanged between the *liquid particle* and the wave-packet, see Figure 2, and has a quantum nature, as argumented above.

The elementary interactions between *liquid particles* and wave packets satisfy the postulate of microscopic time reversibility [119-120]; at the same time, they are at the base of the viscosity model here proposed, representing the building block of a dissipative macroscopic mechanism. As is well known, Loschmidt [134] highlighted the paradox that a thermodynamic system should not irreversibly evolve towards equilibrium if the dynamics of the elementary constituents of the system, either atoms or molecules, is temporally symmetric. This argument was used by Loschmidt to demonstrate the contradiction with the time's arrow of (almost) all the elementary processes with the attempt to deduce from them the second law of thermodynamics, which describes the behaviour of macroscopic systems. Because either the laws of dynamics, that regulates the microscopic elementary processes, or the second law of thermodynamics, that describes the evolution towards equilibrium of macroscopic multi-component systems, are verified, this generates the paradox. Such a paradox was born as a criticism to the so-called *H theorem* by Boltzmann [135] to explain on a statistical base the origin of a macroscopic behaviour of a system starting from the dynamics of its microscopic constituents. Other than Boltzmann, this paradox was discussed and clarified by many authors [see for instance 136-137], up to Onsager with his postulate. In conclusion, although the elementary interactions between the microscopic constituents of liquids are temporally reversible, and provide a physical mechanism for the viscosity in (dual) liquids, that is a dissipative effect, there is no contradiction with the Boltzmann statistical theory of time-asymmetric irreversible equilibrium behaviour.

At the end of Section 3 we have advanced that the physical mechanism proposed for the viscosity in dual liquids is valid also for the friction that originates at a solid-liquid interface. In a previous paper [44], we have not only discussed such topic but also explained how it may give origin to the (faint) thermal gradient observed for the first time by Noirez. The viscosity is one among the scattering processes in which the momentum of normal phonons[1] is dominating and conserved. When phonons are also under a temperature gradient, they can acquire a non-vanishing drift velocity similar to fluid flow driven by a pressure gradient [121]: this is the phonon hydrodynamic regime [41]. The Casimir–Knudsen is one among the heat conduction processes associated with the phonon hydrodynamic regime. In a system we usually distinguish between the phonon mean free path $\langle \Lambda_0 \rangle$ and the characteristic length of the system, say $\langle l \rangle$. A necessary (but not sufficient) condition for the Casimir–Knudsen regime to take place is that $\langle \Lambda_0 \rangle$ is comparable to, or larger than, $\langle l \rangle$. In the configuration of the Noirez experiment, the problem is to consider phonons propagating in the liquid and interacting with the solid plates at the interface, of thickness $\langle l \rangle \pm \delta l$, where $\delta l$ is the uncertain boundary condition. This uncertainty is due to the interface' physical roughness at atomic-scale, responsible for how many phonons are scattered, amounting their wavelength to several nanometers, as estimated also in DML ([41], Table 1). A "specularity" parameter is usually introduced accounting for the fraction of reflected phonons [138]. Therefore, the incoming phonons are responsible for the temperature drop across the interface because of their reflection and transmission. This effect, quantified in Section 4, could explain either the inversion of the temperature drop upon inversion of the plate rotation or the formation of multiple temperature layers detected. A characteristic of the Casimir–Knudsen regime is the linear dependence of thermal conductivity on the specific heat, that is the relation holding for the thermal conductivity $K_l^{wp}$ in the DML (see Eq. (15)). The heat pulse generated during the thermal transient give origin to heat wave propagation (second sound) similar to a mechanical pulse generating a sound wave in a solid, making it possible to describe phonon heat propagation by means of a Cattaneo-type (telegraph) equation [43,121]. In addition to thermal transport, signatures of phonon hydrodynamic transport can be observed from neutron and Brillouin scattering experiments [41,139]. In conclusion, the DML allows a direct comparison with many other interesting aspects characterizing either the solid-like behaviour of liquids, or the presence of phonons as carriers of elastic (and thermal) excitations.

---

[1] In a pure elastic medium, without dissipation, the applied stress and the resulting strain as function of time, such as for instance a momentum flux and the viscous resistance, are exactly in phase, while in a pure viscous medium, they are out of phase by 90°. In a real fluid, the delay lag is in between such limits, and represents a characteristic imprint of the medium.

To conclude, we want to recall the attention of the reader on three aspects, which allow relating the present work with others recently published on the dynamics of liquids at mesoscopic scale. The first concerns the recent paper by Zaccone and Baggioli [118], who proposed a universal law for the vibrational density of states in liquids (VDOS), $g(\omega)$, using the Instantaneous Normal Modes (INMs). This model has allowed for the first time to get an expression for the VDOS reproducing the functional dependence $g(\omega) \propto \omega$ typical of liquids at low energy. We already raised the question on whether there is a relation between the parameter $m$ and $g(\omega)$, as got from [118], or also where and how enter the INMs in the DML? INMs are by definition overdamped oscillatory modes characterized by having only the Im part. Therefore, after having been generated they disappear quite immediately. Are the INMs those modes that originate through the elementary wave-packet↔*liquid particle* interaction? They indeed disappear quite immediately, lasting $\langle \tau_p \rangle$, of the order of few picoseconds (Table 1 in [41]). The second point concerns the comparison of the DML with recent theoretical models on systems exhibiting ***k***-gaps [140]. We have already extensively dealt with such topic elsewhere [41,43], pointing out the several common features with DML, in particular that the presence of ***k***-gaps in liquids may be associated with the fast sound and positive sound dispersion (PSD). Frenkel was the first to point out that a non-zero shear modulus of liquids would have implied a cross-over of the sound propagation velocity from its hydrodynamic value $v = \sqrt{B/\rho}$ to the solid-like elastic value $v = \sqrt{(B + 4/3\,G)/\rho}$, where B and G are bulk and shear moduli, respectively. This topic was taken up again years later by Nettleton [78], who proposed a thermodynamic approach to the viscoelasticity in liquids. Propagation of shear modes occurs at high ***k*** values, implying PSD at these ***k*** points and, further, that PSD should disappear with temperature starting from small ***k*** because the ***k***-gap increases with temperature. Here we want to submit to the reader the question of whether the activation energy for the viscosity in liquids is strictly related to the presence of the ***k***-gap. Such question comes from the fact that, as explained before, the activation energy in the DML should be mostly dependent upon the energy absorbed by the internal collective DoF of the *liquid particle*. Particle dynamics in liquids is characterized either by diffusive jumps, associated with viscosity, into neighbouring locations and enabling liquids to flow, or oscillatory motions at quasi-equilibrium positions, allowing liquids to show a solid-like behaviour [74,82,111,113,38-41, 140]. We have identified the non-linear interaction *liquid particle*↔wave-packet as the dynamics required to allow for both oscillation and jumps activated over potential barrier of the inter-particle potential. This type of interactions are the typical ones dealt with in Nonaffine treatment, allowing to account for the Nonaffine displacements at atomic/molecular level. This argument introduces the third topic, i.e. that of a comparison between the DML and the results of a general theory of collective phonon modes [141], in which field theoretical methods and hydrodynamic are gathered. It is based on Nonaffine displacements and the consequent phase relaxation of collective excitations. Among its consequences there is the presence in liquids of shear waves above a critical value for wave-vector ($k$-gap), assessed on a theoretical firm basis, as well as the theoretical explanation of PSD in liquids for longitudinal waves. One of the key-points of this model is the relaxation phase $\Omega$, that we identify as the inverse of the relaxation time $\langle \tau \rangle$ in the DML. $\Omega$, determining the propagation distance of the collective phonon modes through the exponential decay law $e^{-\Omega \cdot t} \equiv e^{-t/\langle \tau \rangle}$. The conclusion arrived at by the authors is that a liquid may be seen as a system in which the phonon collective modes, those we identify as *lattice particles*, are confined in clusters of size $\langle \Lambda \rangle \approx u^0 \cdot \langle \tau \rangle$, that we identify with the *liquid particles*. At larger distances the system looses the rigidity typical of solids and the associated capability of propagating the mechanical stress by means of shear waves. The reader will have already recognized that this is exactly the picture on which the DML is based, and at the base of the viscosity model described in this manuscript. It is a very surprising and unique circumstance that independent groups have reached similar conclusion on the mesoscopic dynamics of liquids. However, the new ingredient of the DML, consisting in the capability of wave-packets to allow the momentum (and energy) transfer between such clusters, should increase the distance over which the clusters may communicate.

## 6. Conclusions and future perspectives.

In this paper, we have modelled the viscosity of liquids $\eta_l^{wp}$ within the framework of the dual model, providing also the basis for an Arrhenius-like trend vs temperature as well as the comparison of the theoretical result with several experimental situations and with numerical modelling. In particular, the comparison was done demonstrating the successful theoretical interpretation of the mechano-thermal effect revealed by the Noirez' group. This was possible thanks to the duality of liquids, because the wave-packets, travelling between two successive collisions with *liquid particles*, behave as carriers and exchangers of momentum and energy. The successful modelling of $\eta_l^{wp}$, therefore, represents another brick added to the building of the DML.

We have also discussed many interesting insights deriving from a thorough analysis of the consequences of the viscosity modelling. Among these, we want to point out the attention on those that are more susceptible of being investigated from the experimental point of view (other than deepened from the theoretical and numerical simulation ones, of course). The first is related to the crossed effect revealed by Noirez in liquids under shear, that, from our point of view, is not unexpected but rather a natural phenomenon arising in Non Equilibrium Systems. We have demonstrated, in fact, that it may be envisaged as the response of the system to an external stimulus, identified in this case with the momentum gradient. Therefore an interesting experimental campaign could be pursued, dedicated to the detection and to the accurate measurement of this phenomenon in a suitably designed device. Due to the faintness of the thermal gradient and to the particular setup, the microgravitational environment could represent the best environment where to perform such investigation. Thanks to the ever-increasing miniaturization of electronics, an orbiting laboratory the size of Cubesat [142] should already be sufficient to accommodate all the necessary h/w equipment. The experiments could be performed following several setup. The more simple is of course having a single liquid layer of a right thickness, confined in between two solid flat surfaces, with one mobile (or both, counter-rotating, to enhance the effect). The temperature difference can be detected by means of interferometric techniques. A second setup could be that of having two (or more) immiscible liquid layers, one facing the other. This setup is, however, much more difficult to be realized in microgravity because of the fact that surface forces would become dominant in the virtual absence of gravity. Whatever the previous setup is chosen, an external additional thermal source could be added to the device to generate a thermal gradient inside the liquid. We have seen, in fact, that the wave-packets are driven by the internal virtual temperature gradient [41], and that the presence of an additional thermal gradient enhances their flux. Therefore, it is legitimate to await that an external thermal gradient would affect the shear motion of the liquid and, in turn, its response, i.e. the viscosity.

Another intriguing aspect that deserves of being explored is the transient phase of a system after the application of a thermal gradient. For the same reason as above, being the dynamics at mesoscopic level strongly influenced in the DML framework by the presence of a thermal gradient, the more interesting aspects due to the application of such an external stimulus is just the transient phase, that is normally skipped by all experimentalists due to the uncertainties which characterize it, being the local gradients strongly influenced by the external time-dependent conditions. Several experimental suggestions have already been provided in previous papers [41,43]

Finally, the model of parameter $m$ and its dependence upon the main ambient conditions, i.e. the temperature and the pressure, for a chemical species, necessitates of a dedicated theoretical and numerical analysis. A better knowledge of $m$ would provide another strong support to the validity of DML. A first indication is obtained by adaptation of the results of Ghandili et al. [122,126]. We would not be surprised if, at the end, an Arrhenius dependence even for $m$ shall be obtained.

We like to close this manuscript quoting a sentence by Egelstaff [132, chapter 12.6, pag. 159] about the physical origin of the thermal conductivity and of the viscosity of liquids: "*The thermal conductivity of a crystal lattice is related to the scattering of high-frequency phonons. Since the thermal conductivity of a liquid is similar to that of a solid, it is probable that it is due to similar* *phenomena. The results of the ratio* $\eta_l = \dfrac{K_l}{C_l}$ *suggest that the viscosity of a liquid may also be related to the scattering of high-frequency modes of motion".*

**Funding:** This research received no external funding

**Data availability statement:** Not applicable

**Conflicts of interests:** The author declares no conflict of interests.

## 7. Figures

### Figure 1

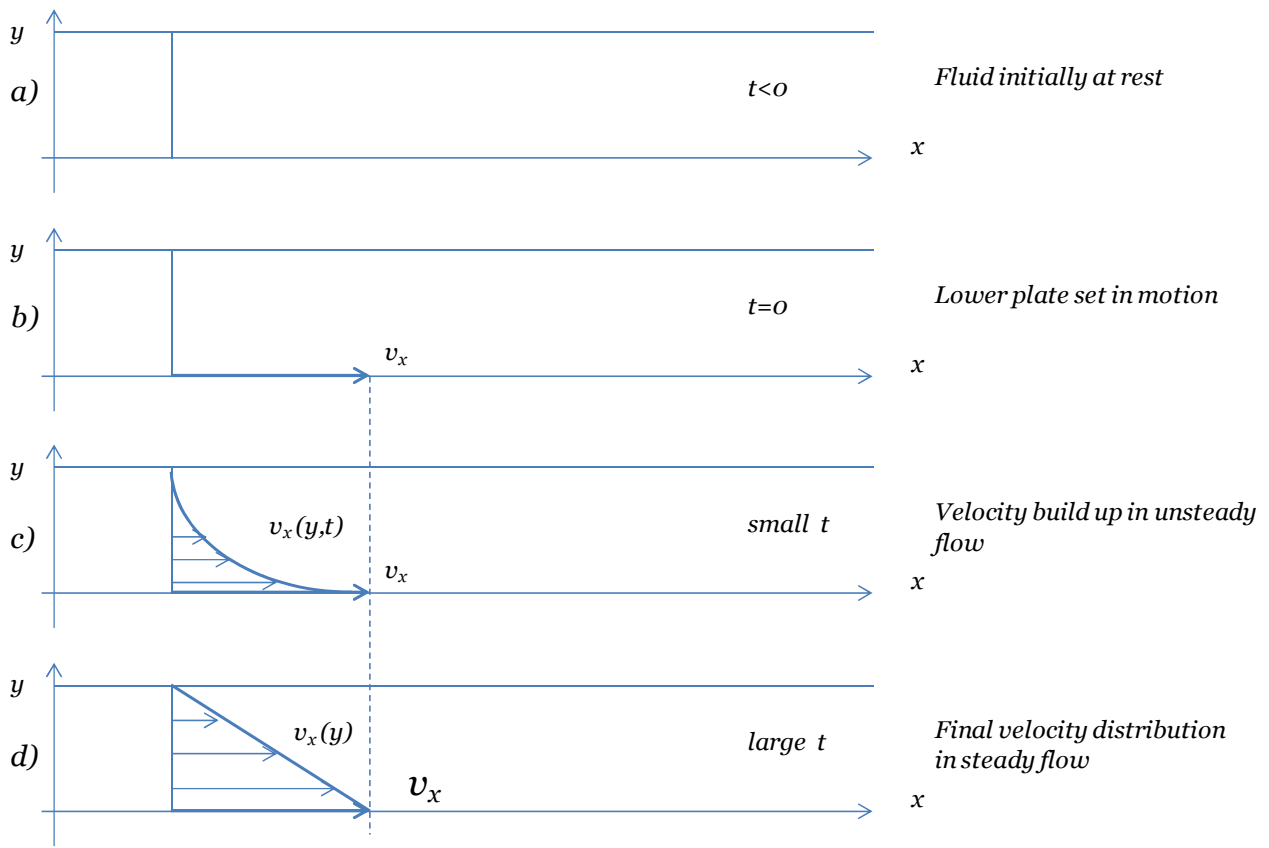

**Figure 1.** The build-up to the steady, laminar velocity profile for a fluid confined between two plates. The flow is called "laminar" because the two adjacent layers of the fluid, i.e. the "*laminae*", slide past one another in an ordered fashion. When the slope of the velocity profile in the last drawing is linear, as in the present case, the liquid is *Newtonian*, and Eq.(1) holds. As usual, an infinite extension is foreseen in the 'z' direction, to avoid sidewall anomalies.

**Figure 2**

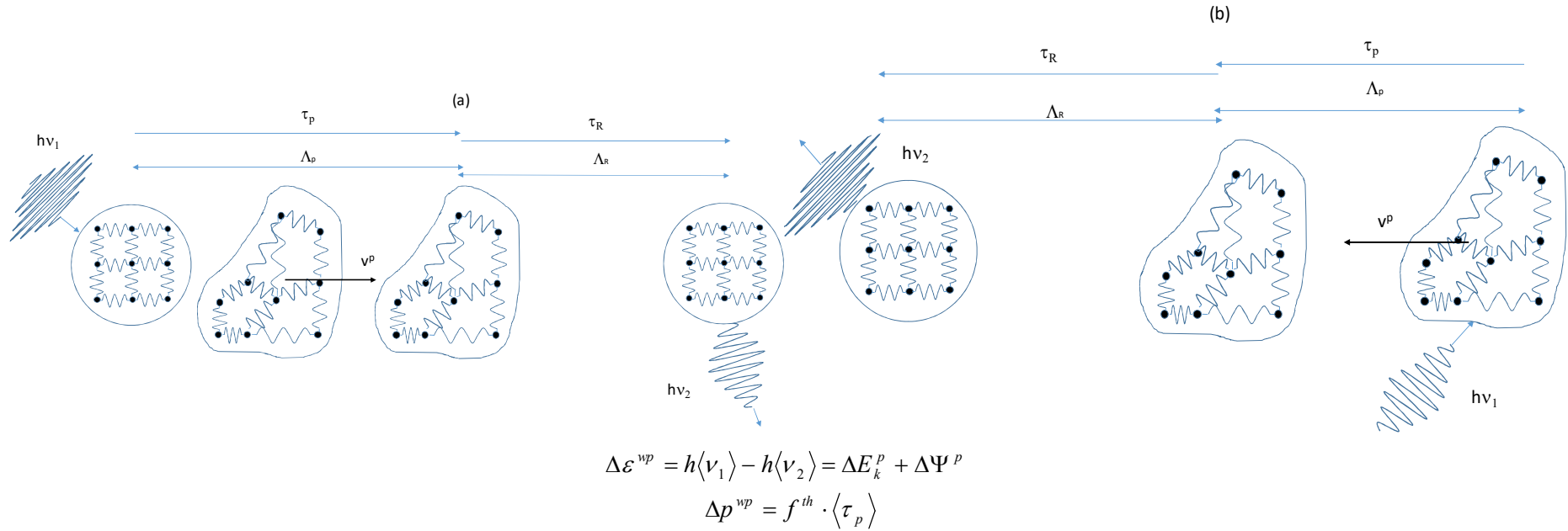

$$\Delta \varepsilon^{wp} = h\langle \nu_1 \rangle - h\langle \nu_2 \rangle = \Delta E_k^p + \Delta \Psi^p$$

$$\Delta p^{wp} = f^{th} \cdot \langle \tau_p \rangle$$

**Figure 2.** Schematic representation of inelastic collisions between wave-packets and *liquid particles*. To make more intuitive the model, we have ideally divided such an elementary interaction in two parts: one in which the wave-packet collides with the liquid particle and transfers to it momentum and energy (both kinetic and potential) and the other in which the *liquid particle* relaxes and the energy is returned to the thermal reservoir through a wave-packet, alike in a tunnel effect (see Figure 4 in [41]), that is a fingerprint of these interactions. In the event represented in (a), an energetic wave-packet transfers energy $\Delta \varepsilon^{wp}$ and momentum $\Delta p^{wp}$ to a *liquid particle*. The particle changes velocity, and the frequency of wave-packet is shifted by the amount $(\nu_2 - \nu_1)$. The interaction is commuted upon time reversal into the one represented in (b), where a *liquid particle* transfers energy and momentum to a wave-packet. This elementary interaction has an activation threshold for potential energy $\Delta \Psi^p$ and, depending on how much energy is absorbed by internal DoF, the rest will become kinetic energy $\Delta E_k^p$ of the liquid particle. The time symmetry of this elementary interaction allows it be assumed the equivalent of Onsager' reciprocity law at microscopic level.

**Figure 3**

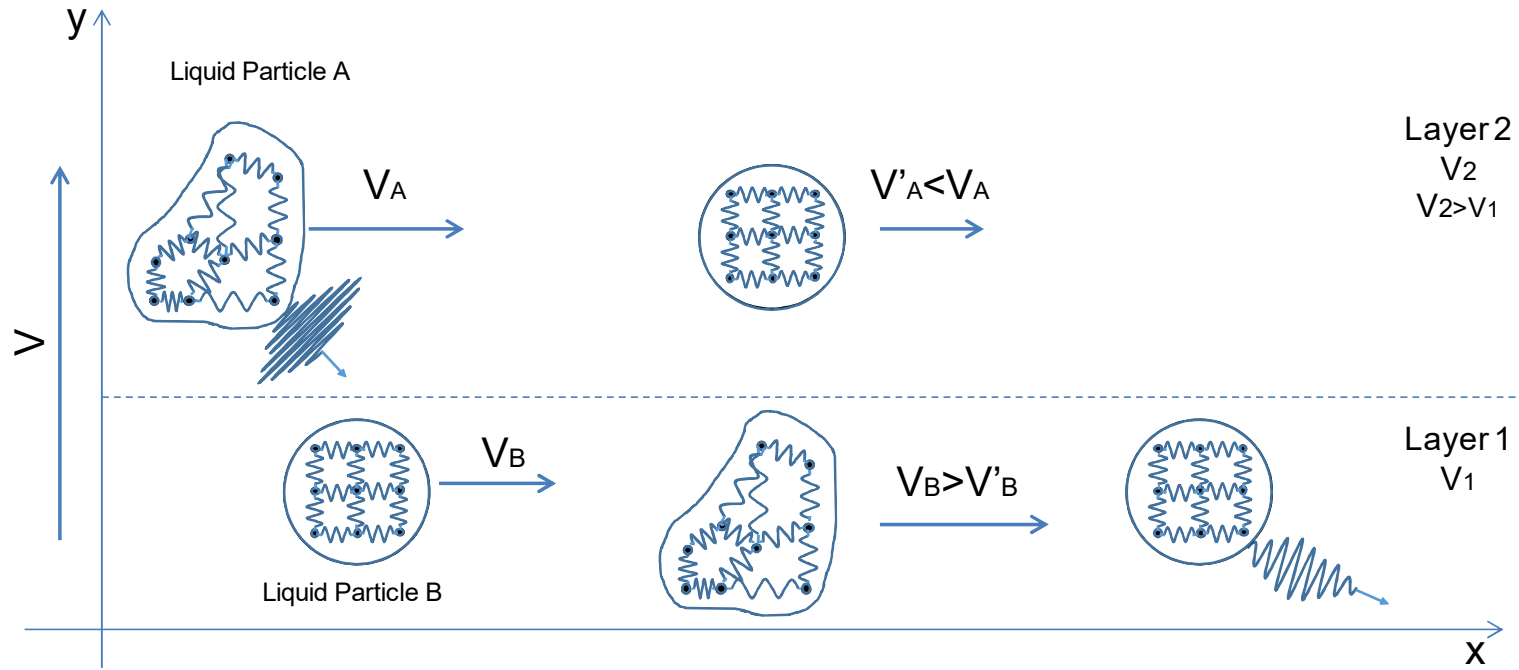

**Figure 3.** Schematic representation of the physical mechanism at the base of the viscosity in the DML. The layer '1' moves at speed $\upsilon_1$ and an adjacent layer '2' moves at speed $\upsilon_2$, with $\upsilon_2 > \upsilon_1$. Being the viscosity represented as the friction between the two layers, this force works like a shear stress, as the fast layer pushes the slow one. In the DML, such friction is mediated by the presence of the wave-packets, responsible for the momentum (and energy) transport pertaining to the solid-like collective DoF. A *liquid particle* A belonging to the Layer #2, moves at speed $\upsilon_2$ through the liquid in the $x$-direction, and a *liquid particle* B belonging to the Layer #1, moves at speed $\upsilon_1$ always in the $x$-direction. The *liquid particle* A is in an excited state and delivers an energetic wave-packet following the process b) of Figure 2; this crosses the border layer and hits the *liquid particle* B. As consequence of the emission, the *liquid particle* A goes in a de-excited state and slows down, due to the recoil. The opposite fate occurs to the *liquid particle* B: it acquires momentum and energy as consequence of the interaction with the wave-packet, as in an event a) of Figure 2. The acquired momentum and energy increase its velocity and the internal energy reservoir, exciting the collective DoF. The kinetic energy will be dissipated by collisions with the other particles of Layer #1, while the energy acquired by the internal DoF will be released into the thermal reservoir of Layer #1.

**Figure 4**

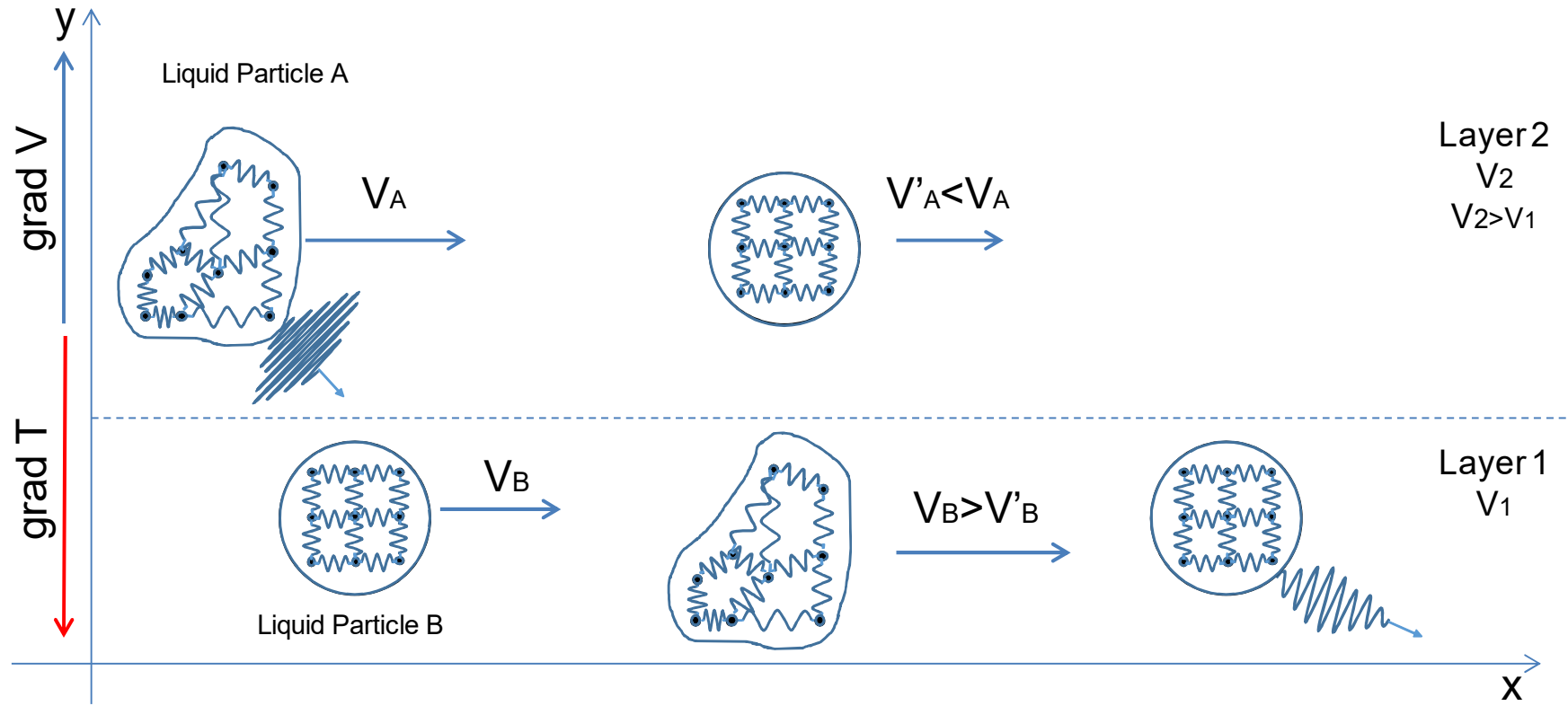

**Figure 4.** Same as Figure 3 but with indicated the external velocity gradient and the induced thermal gradient due to the current of wave-packets.

# 8. Tables

### Table 1

| $\langle \nu_0 \rangle$ | $\langle \Lambda_0 \rangle$ | $\eta_l^{wp}$ |
|---|---|---|
| $0.95 \cdot 10^{12}$ | $1 \cdot 10^{-9}$ | $0.3 \cdot 10^{-3}$ |
| $2.5 \cdot 10^{12}$ | $3 \cdot 10^{-9}$ | $7.5 \cdot 10^{-3}$ |

**Table 1.** Values of dynamic viscosity evaluated in water at ambient conditions of temperature and pressure. All values are expressed in IS units. For both $\langle \nu_0 \rangle$ and $\langle \Lambda_0 \rangle$ we have used minima and maxima values deduced from experiments, as reported in [41]. The value for the density of wave-packets, $\rho_l^{lp}$, is everywhere assumed as $10^3 kg \cdot m^{-3}$.